\documentclass[aps,twocolumn,showpacs,floats,prl]{revtex4}

\usepackage{graphicx}
\usepackage{dcolumn}
\usepackage{bm}
\usepackage{color}
\usepackage{subfigure}

\usepackage[normalem]{ulem}

\begin{document}

\author{ R.~Bombin, J.~Boronat and F.~Mazzanti }

\affiliation{
 Departament de F\'{i}sica, Universitat Polit\`{e}cnica de Catalunya,
 Campus Nord B4-B5, E-08034, Barcelona, Spain }


\title{Dipolar Bose Supersolid Stripes}

\begin{abstract} 
We study the superfluid properties of a system of fully polarized
dipolar bosons moving in the $XY$ plane. We focus on the general case
where the polarization field forms an arbitrary angle $\alpha$ with
respect to the $Z$ axis, while the system is still stable. We use the
diffusion Monte Carlo and the path integral ground state methods to
evaluate the one-body density matrix and the superfluid fractions in
the region of the phase diagram where the system forms
stripes. Despite its oscillatory behavior, the presence of a finite
large-distance asymptotic value in the $s$-wave component of the
one-body density matrix indicates the existence of a Bose
condensate. The superfluid fraction along the stripes direction is
always close to 1, while in the $Y$ direction decreases to a small
value that is nevertheless different from zero. These two facts
confirms that the stripe phase of the dipolar Bose
system is a clear candidate for an intrinsic supersolid without
the presence of defects as described by the Andreev-Lifshitz mechanism.

\end{abstract}

\maketitle 

Supersolid many-body systems appear in nature when two continuous U(1)
symmetries are broken. The first one is associated to the
translational invariance of the crystalline structure, while the
second one corresponds to the appearance of a non-trivial global phase
of the superfluid state~\cite{Boninsegni_2012}. Supersolid phases were
predicted to exist in Helium already in the late
60's~\cite{Andreev_69},
though their experimental observation has been
  ellusive. In fact, the claims for detection made at the beginning of
  this century have been refuted, as the observed behavior is not
  caused by finite non-conventional rotational inertia but rather to
  elastic effects~\cite{Kim_2012}. In this way, a neat observation of
  supersolidity in $^4$He is still lacking. In fact,
it is not clear yet whether a pure, defect-free supersolid
structure like the one that would be expected in $^4$He really exists.
Recently, the issue of supersolidity has emerged again, but now in the
field of ultracold atoms. Two different experimental teams have
claimed that spatial local order and superfluidity have been
simultaneously observed in lattice setups~\cite{Leonard_17} and in
stripe phases~\cite{Li_17}. In this way, the definition of what a
supersolid really is seems to still be under
discussion~\cite{Anderson_2017}.

Superfluid properties of solid-like phases are also of fundamental
interest in quantum condensed matter. One of these is the stripe
phase, where the system presents spatial order in one direction but
not in the others. For instance, stripe phases have been of major
interest since 1990, when non-homogeneous metallic structures with
broken spatial symmetry were found to favor
superconductivity~\cite{Bianconi_00, Bianconi_13}.  More recently,
stripe phases have been observed in Bose-Einstein condensates with
synthetically created spin-orbit coupling~\cite{Li_17},
where the momentum dependence of the interaction induces spatial
ordering along a single direction in some regions of the phase
diagram~\cite{Li_13}. Stripe phases have also been discussed
in the context of quantum dipolar physics, including very recent theoretical 
and experimental analysis of metastable striped gases of $^{164}$Dy~\cite{Wenzel_2017}.
Due to the anisotropic
character of the dipolar interaction, in some regions of the phase
diagram dipoles arrange in stripes, both in
Fermi~\cite{Yamaguchi_10, Sun_10} and Bose~\cite{Macia_2012, Macia_2014}
systems. In some cases the presence of this phase has been reported to
exist even in the isotropic limit~\cite{Parish_12}. Though the
presence of stripe phases in dipolar systems is well established
and has been recently observed~\cite{Kadau_2016}, it
is not yet clear whether the system exhibits superfluid properties
(thus forming
supersolid stripes)
or not.

In a previous work we determined the phase diagram~\cite{Macia_2014}
of the two-dimensional system of Bose dipoles at zero temperature,
tracing the transition lines between the solid, gas and stripe
phases.
The formation and excitation spectrum of the stripe
  phase, where the system acquires crystal order in one direction
  while being fluid on the other,
was previously analyzed in Ref.~{\cite{Macia_2012}}.  In this Letter
we investigate the superfluid properties of the stripe phase as a
function of the density and polarization angle.  Our results show that
dipolar stripes
are a special form of supersolid,
and we quantify the superfluid density
and condensate fraction all along the superstripe phase.

In the following we consider a system of $N$ fully polarized dipolar
bosons of mass $m$ moving on the XY plane. All dipoles are considered
to be aligned along a fixed direction in space given by a polarization
(electric or magnetic) field, which is contained on the XZ plane and
forming and angle $\alpha$ with respect to the Z axis.  The model
Hamiltonian describing the system becomes then
\begin{equation}
H = -{\hbar^2 \over 2m}\sum_{j=1}^N \nabla_j^2
+ {C_{dd} \over 4\pi} \sum_{i<j}^N
\left[ {1 - 3\lambda^2 \cos^2\theta_{ij} \over r_{ij}^3} \right] \ ,
\label{Hamiltonian}
\end{equation}
with $\lambda=\sin\alpha$,
and $(r_{ij},\theta_{ij})$ the polar coordinates associated to the
position vector of particle $j$ with respect to particle $i$.
The constant $C_{dd}$ is proportional to the square of the (electric or
magnetic) dipole moment of the components, assumed all of them to be
identical.  In the following we use dimensionless units obtained from
the characteristic dipolar length $r_0=m C_{dd}/(4\pi\hbar^2)$.

We quantify the superfluid properties of the system
evaluating both the one-body density matrix and its asymptotic value
(the condensate fraction), and the superfluid density. In order to do
that we employ stochastic methods. We use two different quantum Monte Carlo
techniques that are known to provide exact values for the energy of
the system within residual statistical noise: the diffusion Monte
Carlo (DMC)~\cite{Hammond_94, Kosztin_96} and the path integral Monte
Carlo (PIGS)~\cite{Sarsa_2000, Rota_2010} methods. The DMC simulations
have been performed using a second order propagator~\cite{Chin_90},
while a fourth order propagator has been employed in the PIGS
calculations~\cite{Chin_02}. In all cases, a variational model of the
ground state wave function $\Psi_T$
is used.
In the DMC method, the guiding wave function
    is used for importance sampling but the ground state estimation of
    any observable commuting with the Hamiltonian is exact.  In PIGS
    simulations, $\Psi_T$ acts as a boundary condition at the end
    points of the open chains representing the set of particles. It is
    then propagated in imaginary time to the center of the chains,
    where expectation values are evaluated. In this way, any
    contribution orthogonal to the exact ground state is wiped out.
Two different models have been used in this work.
In the DMC simulations, $\Psi_T$
has been taken to be of the Jastrow form, with a two-body correlation
factor that results from the zero-energy solution of the two-body
problem associated to Eq.~(\ref{Hamiltonian}) as derived in
Ref.~\cite{Macia_11}, matched with a long-range phononic extension as
discussed in the same reference.  This model must be modified when
describing the stripe phase, including a one-body term $f_1({\bf r})$
that allows for the formation of the stripes along the $Y$ direction
\begin{equation}
f_1(\textbf{r})=\exp\left[
  \eta_s\cos\left({2\pi n_s y\over L_y}\right)\right] \ ,
\label{onebodystripes}
\end{equation}
with $L_y$ the box side length along the $Y$ direction, and $n_s$ the
number of stripes in the simulation box. Notice that these two
parameters are not independent, as one must guarantee the simulation
box is commensurated for a fixed number of particles.  In
Eq.~(\ref{onebodystripes}), $\eta_s$ is a variational parameter that
is consistently found to be zero in the gas phase, and non-zero in the
stripe phase. For the PIGS simulations,
we have adopted a much simpler
model based on the zero-energy solution of the isotropic ($\alpha=0$)
problem, matched with a phononic tail as in Ref.~\cite{Macia_2014}. Despite
its simplicity, we have found no differences with the results obtained
when using the same model as in the DMC case. 

\begin{figure}
\begin{center}
\includegraphics[width=0.48\textwidth]{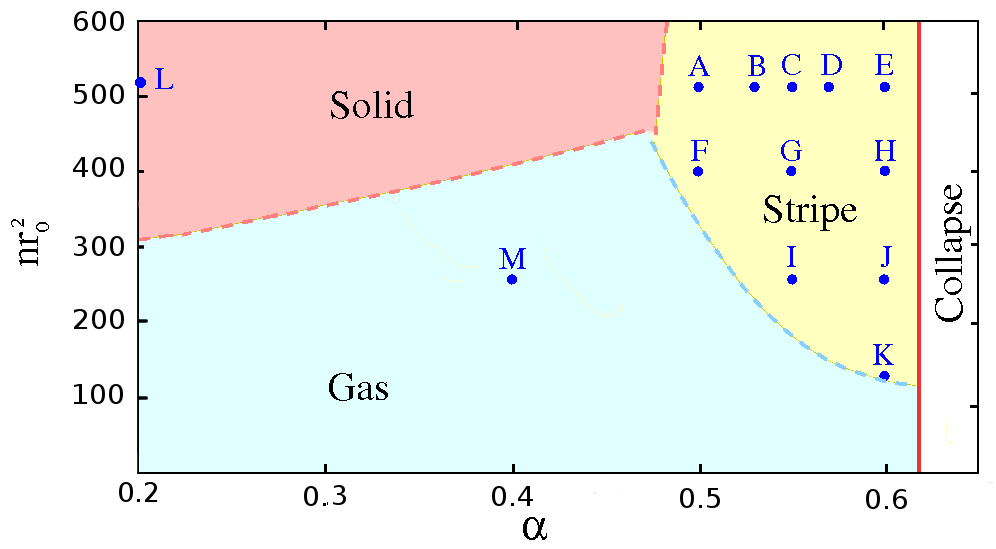}
\caption{(color online). 
Phase diagram of the 2D dipolar Bose gas at zero temperature. 
Letters indicate the set of points corresponding to fixed density and
polarization angles explored in this work.
}
\label{fig_PhaseDiag}
\end{center}
\end{figure}

Since we are analyzing superfluid properties, we have performed
several calculations spanning a wide range of densities and
polarization angles in the
regions of the phase diagram where the system is in stripe
form. Notice that, in the solid phase, the system
    arranges in a triangular lattice that completely breaks the
    continuous translational symmetry~\cite{Macia_2014}, while in the
    stripe phase this symmetry is broken only in one direction (the Y
    axis in our setup).
For the sake of comparison, we have also explored two additional
points where the system remains either as a gas or as a solid. The set
of points explored in this work is shown in the phase diagram,
Fig.~\ref{fig_PhaseDiag}, and a summary of the results obtained for
these points is reported in Table~\ref{table_1}.

\begin{table}[t] 
  \def\arraystretch{1.2}
  \setlength{\tabcolsep}{0.5em}
  \begin{center}
    \begin{tabular}{|c|c|c|c|c|c|c|}\cline{2-7}
      \multicolumn{1}{c|}{} 
      & $n r_0^2$ & $\alpha$ & $n_0$ & $\rho_s$ & $\rho_s^x$ & $\rho_s^y$ \\ \hline
      A & 512 & 0,50 & 0.00030(4) & 0.86(8) & 1.06(8) &
      0.61(8)  \\ 
      B & 512 & 0.53 & 0.00055(6) & 0.62(6) & 0.99(8) & 0.26(3) \\ 
      C & 512 & 0.55 & 0.0029(3) & 0.53(5) & 0.92(8) & 0.14(2) \\ 
      D & 512 & 0.57 & 0.0031(3) & 0.49(5) & 0.95(8) & 0.043(4) \\ 
      E & 512 & 0.60 & 0.0047(5) & 0.49(5) & 0.95(8) & 0.027(3) \\ 
      F & 400 & 0.50 & 0.0038(3) & 1.05(8) & 1.07(8) & 1.04(8) \\ 
      G & 400 & 0.55 & 0.0042(4) &  0.63(6) & 1.001(7) & 0.26(3) \\ 
      H & 400 & 0.60 & 0.0052(4) & 0.55(5) & 1.07(8) & 0.028(3) \\ 
      I & 256 & 0.55 & 0.015(1) & 1.05(8) & 1.03(8) & 1.08(8) \\ 
      J & 256 & 0.60 & 0.011(1) & 0.54(5) & 1.00(8) & 0.080(6) \\ 
      K & 128 & 0.60 & 0.071(4) & 0.95(7) &  0.97(7) &  0.93(7) \\ 
      L & 512 & 0.20 & 0 & 0 & 0 & 0 \\ 
      M & 256 & 0.40 & 0.019(2) & 1 & 1 & 1 \\ \hline
    \end{tabular}
  \end{center}
  \caption{Superfluid densities and condensate fraction for the points shown
  in Fig.~\ref{fig_PhaseDiag}. Figures in parenthesis are the errorbars.}
  \label{table_1}
\end{table}

A direct measure of the off-diagonal long-range order present in the
system is provided by the one-body density matrix (OBDM)
\begin{eqnarray}
  n_1({\bf r}_{11'}) \!\! & = & \!\! 
  \Omega \int d{\bf r}_2\cdots d{\bf r}_N \\
  & & 
  \Psi_0( {\bf r}_1, {\bf r}_2,\ldots,{\bf r}_N)
  \Psi_0( {\bf r}'_1, {\bf r}_2,\ldots,{\bf r}_N) \ ,
  \nonumber
  \label{obdm-b1}
\end{eqnarray}
with $\Psi_0$ the ground state wave function and $\Omega$ the volume
of the container. In this way, $n_1({\bf r})$ is normalized such that
$n_1(0)=1$, while $n_1(|{\bf r}_{11'}|\to\infty)\to n_0$ 
if there is off-diagonal long-range order, with $n_0$
the condensate fraction.
Notice that, in 2D, $n_0$ can be non-zero only at $T=0$.

\begin{figure}
\begin{center}
\includegraphics[width=0.49\textwidth]{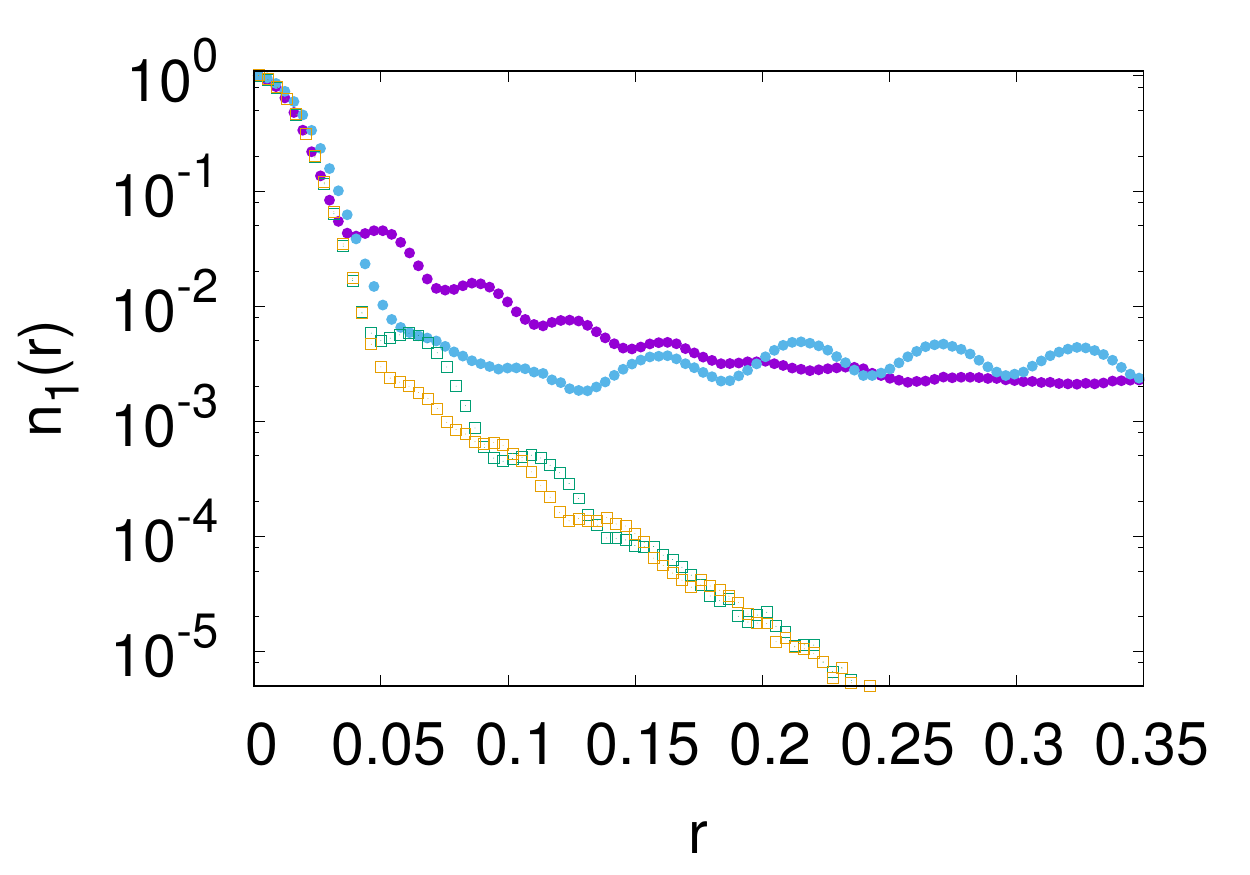}
\caption{(color online).  One Body Density Matrix of the 2D dipolar
  Bose system at the density $n r_0^2=512$ in the stripe phase for
  $\alpha=0.55$ (filled circles), and in the solid phase for
  $\alpha=0.20$ (empty squares).  Purple circles and green squares:
  cuts along the X direction; blue circles and orange squares: cuts
  along the Y direction.  Distance $r$ is measured in units of
  r$_0$. Errorbars are smaller than $10\%$ of each measure and have
  not been included for the sake of clarity.}
\label{fig_OBDM}
\end{center}
\end{figure}

Figure~\ref{fig_OBDM} shows a comparison of the one-body density
matrix of the system at points C and L of Fig.~\ref{fig_PhaseDiag},
corresponding to the same density $nr_0^2=512$ but different polarization
angles. In all cases $n_1({\bf r})$ depends on the direction due to
the anisotropy of the interaction.  The lower curves show two cuts of
$n_1({\bf r})$ along the $X$ and $Y$ directions, when the system is in
the solid phase (point L), while the upper curves show the same
quantities for the system in the stripe phase (point C). As it can be
seen, all curves show an oscillatory behavior that is partially a 
consequence of the anisotropy of the interaction~\cite{Macia_11}. Most
remarkably, the curves corresponding to the solid phase decay
exponentially to zero, while the ones for the stripe phase saturate to
a common value that corresponds to $n_0$. The condensate fraction,
which appears only in the $s$-wave term of the partial wave expansion
of $n_1({\bf r})$, has been obtained by fitting a constant to the
intermediate-distance tail
in regions near (but not at) half the box side where the results are
stable. All values in the third column of Table~\ref{table_1} have
been obtained in this way.

At large densities, where increasing the polarization angle makes the system
change from the solid to the stripe phase, the condensate fraction
increases with increasing $\alpha$. This is not surprising since the
dipolar interaction is overall less repulsive when approaching the
line of collapse, at the critical angle $\alpha_c\approx 0.615$.
The situation is reversed at lower densities, when the system changes from
the gas to the stripe form (point I and J for instance). In this case
and close to the transition line, 
the condensate fraction is expected to approach higher values, as the
gas is less interacting. Perpendicular cuts at fixed polarization
angle and increasing density leads always to a reduction in $n_0$,
consistent with the fact that particles have less effective space.
In any case the largest values of $n_0$ are achieved
near the gas-stripe transition line at the lowest possible densities.
In this way, the large-distance limit of the OBDM of the 
stripe phase is always non-zero, as happens
with other
supersolid systems.

Even though the presence of non-zero condensate fraction value already
points towards a superfluid behavior, it is possible to evaluate 
directly the superfluid response of the system in DMC. 
At finite temperature, the superfluid fraction
  $\rho_s$ is estimated from the winding number~\cite{Pollock_87},
  which takes into accounts the diffusion of world lines at large
  imaginary times.  At $T=0$, this is equivalent~\cite{Zhang_95} to
  measuing the diffusion of the center of mass of the system in the
  infinite imaginary time limit, according to the expression


\begin{equation}
\rho_s = \lim_{\tau\to\infty}
{1\over 4 N \tau} 
\left( {D_s(\tau) \over D_0} \right) \ ,
\label{superflu_b1}
\end{equation}
where $D_s(\tau)=\left\langle ({\bf R}_{CM}(\tau) - {\bf R}_{CM}(0) )^2 \right\rangle$
and $D_0=\hbar^2/(2m)$. For the 2D system analyzed,  
we identify the $X$ and $Y$ components of this expression
with the superfluid fractions
along the $X$ and $Y$ directions, according
to $\rho_s=(\rho_s^x+\rho_s^y)/2$.

\begin{figure}
\begin{center}
\includegraphics[width=0.49\textwidth]{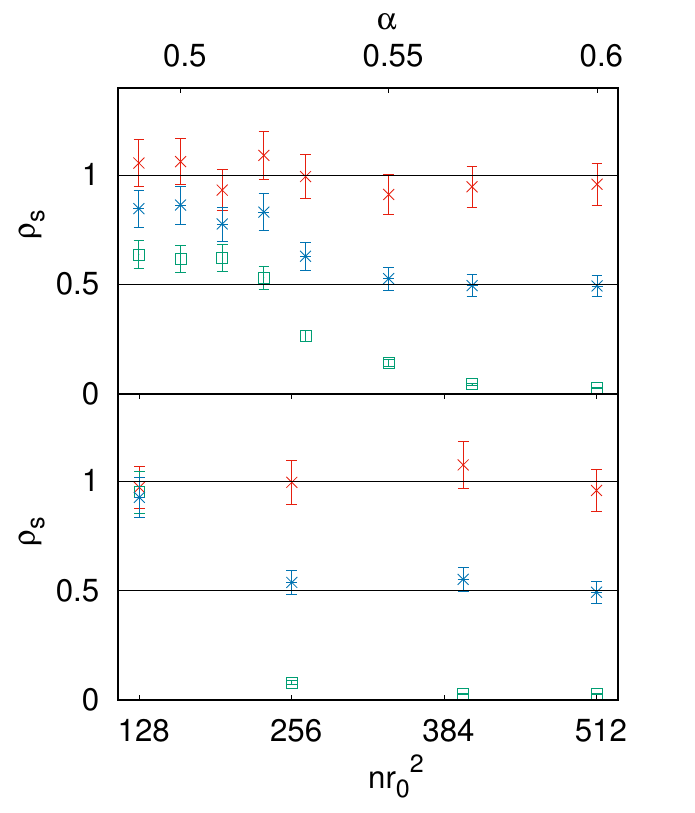}
\caption{(color online).
  Superfluid fractions along the $X$ direction $\rho_s^x$ (red
  crosses), along the $Y$ direction $\rho_s^y$ (green squares) and
  total $\rho_s$ (blue stars). The upper panel shows the dependence of
  these quantities on the polarization angle at the fixed density $n
  r_0^2=512$. The lower panel corresponds to $\alpha=0.6$ and
  different densities. In all cases the system remains in the stripe
  phase. 
  }
\label{fig_rhos}
\end{center}
\end{figure}

Figure~\ref{fig_rhos} shows our results for $\rho_s^x$, $\rho_s^y$
and the total $\rho_s$ for two perpendicular cuts on the phase
diagram. The upper panel corresponds to a fixed density $nr_0^2=512$ and
different angles in the region where the system remains in the stripe
phase. The lower panel corresponds to a fixed angle $\alpha=0.6$ but
different densities, also in the stripe phase. The cut at $nr_0^2=512$ and
increasing $\alpha$ shows that the $X$ component of the superfluid
fraction is always close to $1$, while the $Y$ component decreases to
0, leading to the overall value $\rho_s\approx 1/2$ near
$\alpha=\alpha_c$. Remarkably, the total superfluid fraction $\rho_s$
is larger close to the transition line to the solid phase, decreasing
as $\alpha$ increases. In this way, the superfluid response is
discontinuous across the solid-stripe transition.
The fact that $\rho_s^y$ (and thus $\rho_s$) decrease when $\alpha$
increases is once again a consequence of the anisotropic character of
the dipolar interaction, which becomes less repulsive along the $X$
direction with increasing $\alpha$. Close to $\alpha_c$ the
interaction along the $X$ direction is weak and particles can easily
flow in each stripe, but the confinement of the stripes is stronger
and the system becomes more localized along the $Y$ direction. This is
confirmed by the fact that the optimal values of $\eta_s$ in
Eq.~(\ref{onebodystripes}) is larger when $\alpha$ approaches
$\alpha_c$ at fixed density. A similar situation is found when the
density is increased at constant $\alpha$. The lower panel of
Fig.~\ref{fig_rhos} show the different components of the superfluid
fraction at $\alpha=0.6$ and increasing density. Once again we observe
that $\rho_s^y$ decays to values close to zero already at $nr_0^2=256$,
thus confirming that at high densities the confinement of the
different stripes is very strong. Only point K in that line presents a
large $\rho_s^y$ value, but that point is essentially in the
gas-stripe transition line, and we know the total superfluid fraction
$\rho_s=1$ in the gas phase. Contrarily to what happens when moving
from the stripe to the solid phase, in the gas-stripe transition the
change in $\rho_s$, $\rho_s^x$ and $\rho_s^y$ appears to be continuous.

\begin{figure}
\begin{center}
\includegraphics[width=0.49\textwidth]{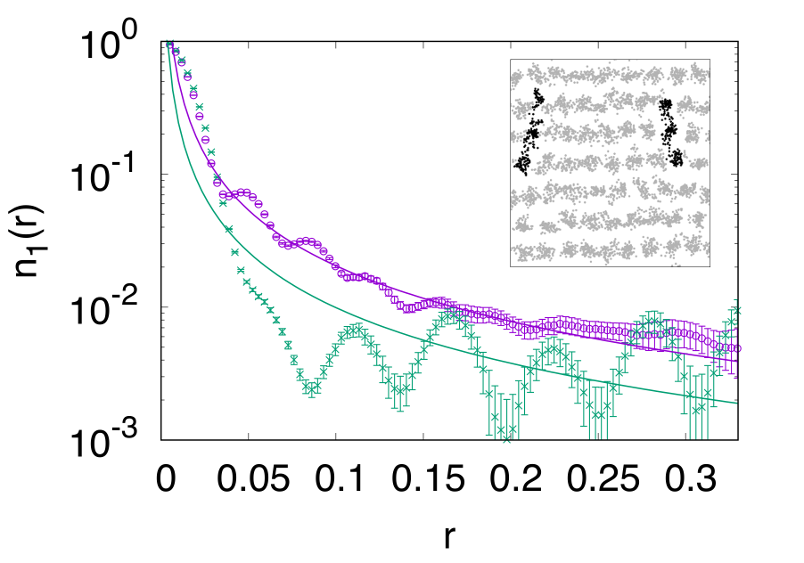}
\caption{(color online).
  One-body density matrix along the $X$ (blue open squares) and $Y$
  (green stars) at $\alpha=0.6$ and $nr_0^2=512$. 
  The solid lines are fits of the form $A x^{-1/\eta}$ with for fixed
  $\eta$ obtained from the slope of the static structure factor near
  the origin. The inset shows a snapshot of the PIGS simulation, where
  some of the particle exchanges are highlighted in black.
  }
\label{fig_n1_Lutt}
\end{center}
\end{figure}

At this point, and according to the previous results, one could wonder
whether stripes are so tightly confined that no particle 
exchanges between different stripes is possible.  If that was the
case, one could also think that each stripe may behave as an isolated,
(quasi) 1D system. In fact and according to the results in the last
column of Table~\ref{table_1}, in some regions the $Y$ component of the superfluid
fraction acquires very low values. However, it never vanishes.  This
indicates that, in fact, particle exchange between different stripes is always
possible, though it becomes unlikely in the limits commented
above. 

Taking that into account, one can look for traces of a (quasi) 1D
behavior in the regions where $\rho_a^y\sim 0$.  One way
to do that is to analyze the system as a Luttinger liquid, and to
check for consistency in the values of the corresponding Luttinger parameters.  
In order to do that we have extracted the sound velocity $c$ from a fit of the 
form $|k|/2c$ to the low-$k$ behavior of the static structure factor $S({\bf k})$ 
evaluated both in DMC and PIGS.
Once with it, we have performed a fit of the form $n_1(u)=A
u^{-1/\eta}$ with $\eta=2\pi n/c$ ~\cite{Luttinger_81} to the $X$ and
$Y$ components of $n_1({\bf r})$, with the results shown in
Fig,~\ref{fig_n1_Lutt}. As it can be seen, the fit reproduces better
the tail of $n_1({\bf r})$ along the $X$ direction, while strong
oscillations in the $Y$ component are clearly visible and $n_1({\bf
  r})$ for ${\bf r}=(0,y)$ differs significantly from the fit.  It
must be kept in mind, though, that the large distance behavior of
$n_1({\bf r})$ in Luttinger liquid theory is a decaying power law not
compatible with a finite condensate fraction value, while we have seen
before that the stripe phase OBDM presents a large-distance asymptotic
value $n_0\neq 0$. In this way, the curve fits well the calculated
$X$-component of the OBDM at intermediate distances only.  The inset
in Fig.~\ref{fig_n1_Lutt} shows a snapshot of the system after
thermalization in PIGS, for the same conditions $nr_0^2=512$ and
$\alpha=0.6$, where a pair of examples where particle exchange between
different stripes is visible, have been highlighted. It is worth
recalling that since simulations in PIGS are done with open chains
(with variational wave functions at the end points), it is hardly
possible to see long exchange lines crossing the whole simulation box.

In summary, we have performed DMC and PIGS simulations to analyze the
supersolid
properties of
dipolar Bose
stripes in
two dimensions for
polarization angles before collapse. We have evaluated the one-body
density matrix to find that it always presents a finite (though in
some regions, quite small) condensate fraction value, in contrast to
the continuously decaying tail it presents in the solid phase. We have
also evaluated the superfluid fraction along the $X$ and $Y$
directions to find that, at large densities and/or polarization angles,
the $Y$-component becomes very small, though it never vanishes. At
high densities and polarization angles the stripes are tightly
confined and the intermediate distance behavior of the OBDM along the
stripe direction has a dependence on the distance that is somehow compatible
with a Luttinger liquid model. However, particle exchanges, always
visible in configuration snapshots, lead to a finite condensate
fraction value and an overall superfluid behavior that, together with
the existence of Bragg peaks~\cite{Macia_2014}, confirm the
supersolid
character of that phase.

This work has been supported by the Ministerio de Economia, Industria
y Competitividad (MINECO, Spain) under grant No. FIS2014-56257-C2-1-P.

\end{document}